\documentclass[10pt,conference]{IEEEtran}
\newif\ifanonymous
\anonymousfalse %
\usepackage{amsmath,amssymb,amsfonts}
\usepackage[utf8]{inputenc} %
\usepackage{booktabs} %
\usepackage{array}    %
\usepackage{tabularx} %
\usepackage[hidelinks]{hyperref}
\usepackage[nameinlink,capitalise,noabbrev]{cleveref}
\crefname{lstlisting}{Listing}{Listings} %
\Crefname{lstlisting}{Listing}{Listings}
\usepackage{multirow}
\usepackage{threeparttable}
\usepackage{adjustbox}
\usepackage{diagbox}
\usepackage{makecell}
\usepackage[inline]{enumitem} %
\newlist{inlist}{enumerate*}{1}
\setlist[inlist]{itemjoin={{, }},itemjoin*={{, and }},label=($\roman*$),mode=boxed}
\newlist{inlistAsWell}{enumerate*}{1}
\setlist[inlistAsWell]{itemjoin={{, }},itemjoin*={{, as well as }},label=($\roman*$),mode=boxed}
\usepackage{csquotes} %
\let\say\enquote
\usepackage{textcomp} %
\usepackage{listings} %
\usepackage{xcolor}
\usepackage{caption}
\usepackage[colorinlistoftodos]{todonotes}
\newcommand{\commentToReviewer}[2][]{\todo[#1]{#2}} %
\renewcommand{\commentToReviewer}[2][]{}
\usepackage[detect-all]{siunitx}
\PassOptionsToPackage{protrusion=true, activate={true,nocompatibility}, final, tracking=true, kerning=true, spacing=true, factor=1100}{microtype}
\usepackage{etoolbox}  %
\usepackage{microtype}
\usepackage{xspace} %
\usepackage{xparse}
\def\FMeasure{F\textsubscript{1}\xspace}
\newcommand{\dash}{\thinspace---\thinspace\xspace} %
\usepackage{orcidlink} %
\usepackage[subtle]{savetrees} %
\usepackage[most]{tcolorbox}
\newtcolorbox{AnswerBox2}{
  breakable,
  enhanced,
  colback=white,
  colframe=black,
  boxrule=0.5pt,
  arc=6pt,         %
  outer arc=6pt,   %
  boxsep=0pt, %
  left=2pt,
  right=2pt,
  top=2pt,
  bottom=2pt,
  before skip=6pt,
  after skip=6pt,
}

\DeclareRobustCommand\HypothesisLabel[1]{\ifcase#1
\or $H_{01}$%
\or $H_{A1}$%
\or $H_{02}$%
\or $H_{A2}$%
\else ???\fi}

\DeclareRobustCommand{\Standalone}{\textit{Neumüller et al.}\xspace} %

\newlength{\belowTableSaveSpace}
\begin{document}
\raggedbottom %
\title{Combining Static Code Analysis and Large Language Models Improves Correctness and Performance of Algorithm Recognition}

\author{
\IEEEauthorblockN{Denis Neumüller\,\orcidlink{0000-0003-3872-0188}}
\IEEEauthorblockA{\textit{Ulm University}\\
Germany \\
denis.neumueller@uni-ulm.de}
\and
\IEEEauthorblockN{Sebastian Boll\,\orcidlink{0009-0004-7618-2469}}
\IEEEauthorblockA{\textit{Ulm University}\\
Germany \\
sebastian.boll@uni-ulm.de}
\and
\IEEEauthorblockN{David Schüler\,\orcidlink{0009-0000-7665-9707}}
\IEEEauthorblockA{\textit{Ulm University}\\
Germany \\
david.schueler@uni-ulm.de}
\and
\IEEEauthorblockN{Matthias Tichy\,\orcidlink{0000-0002-9067-3748}}
\IEEEauthorblockA{\textit{Ulm University}\\
Germany \\
matthias.tichy@uni-ulm.de}
}

\maketitle

\begin{abstract}
\textit{Context:}
Since it is well-established that developers spend a substantial portion of their time understanding source code, the ability to automatically identify algorithms within source code presents a valuable opportunity.
This capability can support program comprehension, facilitate maintenance, and enhance overall software quality.
\textit{Objective:}
We empirically evaluate how combining LLMs with static code analysis can improve the automated recognition of algorithms, while also evaluating their standalone performance and dependence on identifier names.
\textit{Method:}
We perform multiple experiments evaluating the combination of LLMs with static analysis using different filter patterns.
We compare this combined approach against their standalone performance under various prompting strategies and investigate the impact of systematic identifier obfuscation on classification performance and runtime.
\textit{Results:}
The combination of LLMs with lightweight static analysis performs surprisingly well, reducing required LLM calls by 72.39--97.50\% depending on the filter pattern.
This not only lowers runtime significantly but also improves F1-scores by up to 12 percentage points (pp) compared to the baseline.
Regarding the different prompting strategies, in-context learning with two examples provides an effective trade-off between classification performance and runtime efficiency, achieving F1-scores of 75--77\% with only a modest increase in inference time.
Lastly, we find that LLMs are not solely dependent on name-information as they are still able to identify most algorithm implementations when identifiers are obfuscated.
\textit{Conclusion:}
By combining LLMs with static analysis, we achieve substantial reductions in runtime while simultaneously improving F1-scores, underscoring the value of a hybrid approach.
\end{abstract}


\section{Introduction}\label{sec:Introduction} %
It has long been shown that developers spend more than half of their time with program comprehension~\cite{Xia2018, Minelli2015}.
Naturally, many different approaches have been developed to support this activity such as the recovery of software architecture design decisions~\cite{DBLP:conf/icsa/ShahbazianLLBM18}, the recovery of models~\cite{DBLP:journals/access/RaibuletFZ17}, or the automatic detection of algorithms~\cite{Kozaczynski1992, Quilici1994, Wills1994, Metzger2000, Alias2003, Zhu2011, Taherkhani2013, Mesnard2016, Nunez2017, ARCC2, Long2022, Neumueller2024}. %
Approaches from the last area focus on automatically detecting algorithms in source code bases with the goal of supporting various software engineering activities such as supporting program comprehension, improving software quality, and enabling advanced analysis or optimizations.
An \say{algorithm} in this context can be defined as \say{a specific computational procedure~[\ldots] for solving a well-specified computational problem} following Cormen et al.~\cite[p. 5]{DBLP:books/daglib/0023376}, e.g., sorting, handling data structures, computing shortest paths in a graph, etc.

In a reverse-engineering context, developers often approach a code base or component with which they have little prior familiarity.
Detecting which algorithms are implemented in the code base helps them understand which problem the code addresses, how the concern is solved — i.e. which specific algorithm implementation is used — and which components of the source code base are involved in the implementation.

For example, detecting the implementation of Dijkstra's Algorithm allows developers to immediately infer that the code deals with graph-based structures and path optimization e.g. for route planning or dependency resolution in build automation tools.
Detecting the implementation of the Raft consensus algorithm signals that the system is designed to ensure fault-tolerant, distributed state replication, used for example in distributed databases or container orchestration services.

Additionally, a computational problem can be solved by multiple different algorithms with different characteristics (e.g. time and space complexity, guarantees about the optimality of the solution, \ldots).
Knowing which concrete algorithm is used in the system also gives developers insights into the design decisions of the original developers as well as the system's quality.

A recent user study by Neumüller et al.~\cite{Neumueller2025} provides further evidence that providing information about implemented algorithms improves program comprehension.
They also surveyed participants, including professional software developers, and collected over 15 use cases in which algorithm recognition would be considered useful.
One example of the collected use cases include the automatic recognition (and replacement) of suboptimal algorithm implementations.
Such a capability could for example be introduced into the merge request review of a development pipeline, where such optimization suggestions can be presented to the developer along with other linter warnings.
Another mentioned use case of algorithm recognition is its application in teaching assistance tools in order to reduce the reviewing effort for student submissions as well as providing timely feedback to students in the context of university courses.

Since determining whether two programs are semantically equivalent — and therefore algorithm recognition — is impossible in the general case~\cite{Metzger2000}, all existing techniques adopt heuristic or best-effort strategies.
These typically rely on search-patterns, that describe the algorithms to find, or machine learning based approaches which train classifiers for detection.
Depending on the actual implementations the source code is represented as simple text, AST, data-flow, control-flow or a combination of the before.

In the recent past, Large Language Models (LLMs) have show remarkable performance in a variety of code related tasks~\cite{Hou2024, Jiang2024, Zheng2024}, such as code summarization~\cite{2022LLMsCodeSummarization, 2023transformersCodeSummarization}, code translation~\cite{Yang2024, Zhang2025} or bug fixing~\cite{2023APR_LLMs, 2023RepairIsNearlyGeneration}.
However, our paper is the first to comprehensively evaluate a hybrid approach that combines static code analysis with Large Language Models for the task of algorithm detection.
The goal of our paper can be summarized using the GQM template~\cite{Jedlitschka2005} as follows:
\textbf{Analyze} the combination of static analysis and Large Language Models
\textbf{for the purpose of} algorithm detection
\textbf{with respect to} performance and efficiency
\textbf{from the point of view of} researchers and developers
\textbf{in the context of} real-world open source Java code.

From this goal we derive the following research questions:
\begin{enumerate}[label={\textbf{RQ\arabic*:}},ref={RQ\arabic*},leftmargin=*]
  \item \label{rq:1}Which prompting strategies work best for algorithm recognition?
  \item \label{rq:2}Can filter patterns be used effectively to decrease runtime requirements? 
  \item \label{rq:3}How important is name-information for the classification performance of the LLMs?
\end{enumerate}
Addressing \ref{rq:1}, we evaluated various LLMs and prompting strategies on the BCEval dataset, finding that more elaborate prompts improve detection but increase runtime.
Consequently, \ref{rq:2} focuses on optimizing runtime efficiency by combining LLMs with static analysis to exclude as many true negatives as possible before invoking LLMs. 
Lastly, \ref{rq:3} measures the effect of systematic identifier obfuscation on LLM classification performance to quantify their dependence on name information.
To answer \ref{rq:3}, we measure how systematic identifier obfuscation affects LLM classification performance enabling us to quantitatively assess their reliance on name information.

The contributions of our paper include:
\begin{inlist}
	\item the evaluation of different prompting strategies and LLMs for algorithm detection
	\item the combination of LLMs with suitable filter patterns to address performance issues and improve recognition performance
	\item an investigation of how name information influences the detection performance of LLMs.
\end{inlist}

\section{Experimental Design}
We present our study based on the guidelines for reporting empirical studies in Software Engineering outlined by Jedlitschka et al.~\cite{Jedlitschka2005}.
Our design and setup follows the recommendations by Field et al.~\cite{Field2003}, who outline best practices for designing experiments.
Specifically, we employ a repeated-measures design, exposing each LLM to all experimental conditions~\cite{Field2003}.

\subsection{Goals}
As described in \cref{sec:Introduction}, the goal of our experiment was to evaluate the performance of LLMs for algorithms detection tasks.
To investigate this, we evaluated multiple state-of-the-art LLMs (see Sec. \ref{sec:Models}) on the BigCloneEval dataset (see Sec. \ref{sec:Dataset}).
In \cref{sec:RQ1} we focus on different prompting styles and their influence on classification performance and runtime.
Since more complex prompts also increase the runtime of the models \cref{sec:RQ2} investigates the use of filter-heuristics to improve the runtime.
To quantify how strongly the LLMs rely on name information present in the source code, we perform an ablation study obfuscating identifiers and report its results in \cref{sec:RQ3}.

\subsection{Models}\label{sec:Models}
We selected recent models with strong results on code and instruction-following benchmarks, particularly HumanEval~\cite{Chen2021} and IFEval~\cite{Zhou2023}, as summarzied in \Cref{tab:models}.
\begin{table}[h]
    \vspace{-2mm}
    \small
    \centering
    \begin{adjustbox}{width=0.95\linewidth,center} %
    \renewcommand{\arraystretch}{1.3} %
    \begin{threeparttable}
        \begin{tabular}{@{}l>{\centering\arraybackslash}p{1.5cm}>{\centering\arraybackslash}p{1.2cm}>{\centering\arraybackslash}p{1.1cm}>{\centering\arraybackslash}p{1.1cm}@{}}
            \toprule
            \textbf{Model} & \multicolumn{1}{l}{\makecell[c]{\textbf{Active}\\\textbf{Parameters}}} & \makecell[c]{\textbf{Context}\\\textbf{Size}} & \makecell[c]{\textbf{Human}\\\textbf{Eval}\tnote{1}} & \makecell[c]{\textbf{IF}\\\textbf{Eval}} \\
            \midrule
            GPT-4o mini & undisclosed\tnote{2}   & 128k\tnote{3} & \textsuperscript{\phantom{*}}88\%\textsuperscript{\phantom{*}} & 78\%\tnote{4}  \\
            \addlinespace[1mm]
            \makecell[l]{Llama3.1 70B\\ Instruct} & 70B\tnote{5}   & 128k\tnote{5} & \textsuperscript{\phantom{*}}81\%\textsuperscript{*} & 88\%\tnote{5}  \\
            \addlinespace[1mm]
            \makecell[l]{Mixtral 8x22B\\ Instruct} & 39B\tnote{6}   & \phantom{1}64k\tnote{6}  & \textsuperscript{\phantom{*}}72\%\textsuperscript{*} & 72\%\tnote{7}  \\
            \bottomrule
        \end{tabular}
        \begin{tablenotes}
          \footnotesize
          \item[] Sources: \textsuperscript{1}\cite{artificialanalysis_models},
          \textsuperscript{2}\cite{openai2024gpt4technicalreport},
          \textsuperscript{3}\cite{OpenAIGPT4ominiModelApIOverview},
          \textsuperscript{4}\cite{OpenAIGPT4_1},
          \textsuperscript{5}\cite{Grattafiori2024Llama3Herd},
          \textsuperscript{6}\cite{mixtral2024BlogMixtral8x22b},
          \textsuperscript{7}\cite{open_llm_leaderboard};
          \textsuperscript{*}Score for the base model since the instruction-tuned scores are not available.
        \end{tablenotes}
    \end{threeparttable}
    \end{adjustbox}
    \vspace{-1mm}
    \caption{Comparison of selected LLMs.}
    \label{tab:models}
    \vspace*{-3mm}
\end{table}

\textbf{GPT-4o mini} is a state-of-the-art model with good performance in coding and instruction following~\cite{openai2024gpt4technicalreport, OpenAIGPT4omini, OpenAIGPT4_1}.
GPT-4o mini is the only closed-source model in our comparison and was made available on July 2024.
We select the mini variant for its substantially lower cost and near-equivalent performance (3~pp lower on HumanEval), while surpassing GPT-4 in prior user preference benchmarks~\cite{OpenAIGPT4omini}.
Specifically we use the gpt-4o-mini-2024-07-18 snapshot, through OpenAI's Batch API.

\textbf{Llama 3.1 Instruct} is a state-of-the-art instruction-tuned model from Meta's open-source Llama 3 family of models.
Trained on large public datasets these models support multilinguality, coding, reasoning, as well as tool usage and can surpass much larger models in reasoning and coding performance~\cite{Touvron2023Llama, Grattafiori2024Llama3Herd}.
We selected the Llama 3.1 Instruct 70B variant, which is finetuned to follow instructions, shows good performance in coding tasks and is the largest model runnable with 188GB of VRAM.

\textbf{Mixtral 8x22B Instruct v0.1 Q8} is built on Mistral AI's mixture-of-experts architecture~\cite{Jiang2024MixtralPaperSmallModels, mixtral2024BlogMixtral8x22b}.
It comprises eight 22B-parameter experts, but only two are active per inference step, enabling faster inference despite the large model size. 
We use the 8-bit quantized 8x22B Instruct version release in April 2024~\cite{Jiang2024MixtralPaperSmallModels, mixtral2024BlogMixtral8x22b}, to fit our VRAM budget.
For brevity, models will be referred to by their shorthands throughout the paper.

\subsection{Dataset}\label{sec:Dataset}
We perform our experiments using the BigCloneEval (BCEval) dataset~\cite{svajlenko2016bigcloneeval} which contains manually verified algorithm implementations from real-world source code bases.
BCEval was created by Svajlenko et al. who defined 43 different functionalities such as \textit{Web Download}, \textit{Send E-Mail} or \textit{Copy File}.
For each of those functionalities they created a regex heuristic to find candidate implementations in the IJaDataset 2.0~\cite{ija}, which consists of 365 MLOC of Java code from real world open source code projects.
The resulting candidate methods were then labeled by a set of judges as true positives that actually implement the defined functionality and false positives that do not. %
BCEval consists of only the subset of files that contain labeled methods~\cite{svajlenko2016bigcloneeval}. %

Since BCEval was originally created to evaluate code clone detection tools not all the functionalities would typically be considered to be an algorithm.
We therefore select a subset consisting of \textit{Prime Factors}, \textit{Greatest Common Divisor (GCD)}, \textit{Fibonacci}, \textit{Palindrome}, \textit{Bubble Sort}, \textit{Binary Search} and \textit{Transpose Matrix} and use the high quality ground truth to perform our algorithm detection experiments.
We formalize our algorithm detection experiments as a classification problem and iterate over each method in the respective functionality prompting the LLM to decide if the method implements the algorithm in question.

Although datasets sourced from programming contest platforms~\cite{Mou2016, Long2022} provide explicit algorithmic labels, they contain isolated implementations with little surrounding logic and no genuine negative class.
Each sample corresponds to a predefined algorithmic task, resulting in an overly simplified classification setting where detection methods can rely on superficial cues (e.g., the presence of matrix operations) rather than actual algorithmic reasoning.
In contrast, BCEval is derived from real source code bases, where algorithmic logic is partially embedded within broader business functionality. 
Its manually verified labels ensure a high-quality ground truth, and the extensive negative class substantially increases task difficulty, making BCEval a more rigorous and realistic benchmark for evaluating genuine algorithm recognition capability.

Since we only use pre-trained models and do not train any models ourselves we do not require a training split.
However, we do divide the dataset into a test and validation set using a 70:30 ratio, because we manually fine-tune the filter patterns used in \ref{rq:2} (\cref{sec:RQ2}) on the test split.   
We performed the two-sample Kolmogorov-Smirnov test to validate that our created splits are representative of the whole dataset in terms of the number of abstract syntax tree elements per method.
\Cref{tab:dataset_overview} gives an overview of our dataset.

Some of the experiments require excessive runtimes exceeding our compute budget.
The size of BCEval is largely driven by a substantial amount of true negatives in the Bubble Sort and Binary Search functionalities.
We therefore create a \textit{reduced dataset} which is a copy of the full dataset with the same split ratio, but only includes 10\% of the true negatives for these two functionalities.
We again validated its representativeness using the Kolmogorov-Smirnov test.

BCEval's focus is on evaluating recall and it is therefore usually not used to evaluate precision~\cite{svajlenko2016bigcloneeval, Farmahinifarahani2019, SvajlenkoSurveyCCDBenchmarking}.
Since the IJaDataset is mined using regex heuristics it is possible that some methods implementing the functionality were not found and therefore not labeled~\cite{Svajlenko2014}.
This means that some of the \enquote{false positives} reported by a classifier could actually be valid implementations with a missing label. 
It is however possible to calculate a lower bound for the precision by assuming that each detected method unknown to the benchmark is a false positive~\cite{Svajlenko2014}. 
Because precision is a lower bound, the F1-score is also lower bound.
We have adopted this strategy for our evaluation.

Given the imbalance in the number of methods per functionality, we employ the macro-averaged F1-score\footnote{${Macro-avg.\;\;\; \FMeasure-score} = \tfrac{F_{1_{Prime Factors}}\:\: +\:\: \hdots\:\: +\:\: F_{1_{Transpose Matrix}}}{7}$} to prevent overemphasizing the importance of any single algorithm. 
We consider this choice appropriate, as it avoids making assumptions about the distribution of algorithm implementations that may appear in users' codebases.
Furthermore, our objective is to assess the general effectiveness of LLMs for algorithm detection, independent of specific algorithm frequencies.
{
\setlength{\tabcolsep}{4.7pt} %
\begin{table}[h]
    \vspace{-2mm}
    \small
    \centering
    \begin{adjustbox}{width=0.95\linewidth,center}
    \renewcommand{\arraystretch}{1.0}
    \sisetup{
        table-align-text-post=false,
        detect-weight=true,
        mode=text,
        group-separator={,},
        table-number-alignment=center
    }
    \begin{tabular}{l|S[table-format=4.0]S[table-format=5.0]c|S[table-format=4.0]S[table-format=5.0]c}
          & \multicolumn{2}{c}{\textbf{Test Split}} &  & \multicolumn{2}{c}{\textbf{Validation Split}} &  \\ 
    \multicolumn{1}{l|}{}     & {Positive}      & {Negative}       &  & {Positive}        & {Negative}         &  \\ \midrule
    \multicolumn{1}{l|}{\makecell[l]{Prime Factors}} & 15      & 897       &  & 6         & 384         &  \\
    \multicolumn{1}{l|}{\makecell[l]{GCD}} & 16      & 1401      &  & 7         & 600         &  \\
    \multicolumn{1}{l|}{Bubble Sort}   & 117     & 16301     &  & 50        & 6986        &  \\
    \multicolumn{1}{l|}{Palindrome}    & 118     & 2497      &  & 50        & 1070        &  \\
    \multicolumn{1}{l|}{Fibonacci}     & 148     & 2377      &  & 64        & 1019        &  \\
    \multicolumn{1}{l|}{Binary Search} & 282     & 40241     &  & 121       & 17246       &  \\
    \multicolumn{1}{l|}{Transpose Matrix} & 343     & 8784      &  & 147       & 3764        &  \\ \midrule
    \multicolumn{1}{l|}{Total} & \multicolumn{2}{c}{72,537} &  & \multicolumn{2}{c}{31,514}     & 
    \end{tabular}
  \end{adjustbox} %
    \vspace{-1mm}
    \caption{Number of methods in the full dataset.}
    \label{tab:dataset_overview}
    \vspace*{\belowTableSaveSpace}
\end{table}
}

\subsection{Hardware}
The experiments were conducted on machines with two Intel Xeon Gold 6248 CPUs (32 cores, 2.6 GHz), 512 GB RAM and four NVIDIA H100 GPUs, each with 94 GB of VRAM. 
We parallelized runs by executing two LLM instances per node, each allocated to two GPUs (188 GB VRAM per model).

\section{\ref{rq:1}: Prompting Strategies}\label{sec:RQ1}
Apart from the model itself the choice of the \enquote{right} prompt is probably the most important aspect of using LLMs effectively.
Multiple studies show that the performance of LLMs can be increased by modifying the used prompt~\cite{Sahoo2025, Marvin2024, Chen2025}.
In this research question we therefore evaluate different prompting styles as well as providing the LLM with different contextual information to determine the most effective approach for our specific use case.
In the following we first discuss the selection of our baseline before evaluating \textit{in-context learning} as well as \textit{chain of thought} prompting.

\subsection{Baseline}\label{sec:baseline}
To establish a baseline for evaluating prompting techniques, we created a very simple prompt that looked as follows:
{
\vspace{-1mm}
\begin{lstlisting}[stringstyle={\textit},escapeinside=!!]
!\textbf{\mbox{User Prompt}}!: "SNIPPET: !\color{gray}\{method\}! Does the snippet implement !\color{gray}\{algorithm\}!, only answer with 'Yes' or 'No'?"
\end{lstlisting}
\vspace{-1mm}
}

However, depending on the specific use case for which the algorithm detection system is designed, it may be desirable to prioritize precision over recall, or vice versa. 
For example, when identifying suboptimal or low-quality algorithm implementations during automated code review, precision is critical to avoid overwhelming users with false positives.
Conversely, when detecting potentially unlicensed or proprietary algorithm implementations for the purpose of intellectual property protection, recall may be more important. 
Missing even a single instance of unauthorized usage may lead to legal, financial, or reputational risks.
To facilitate different use cases we also designed a \textit{score} prompt.
Instead of a binary yes or no answer we instruct the LLM to rate each provided method with a score from 0\,--\,4 representing how likely the code implements the algorithm in question.\footnote{We also tested a finer 0\,--\,10 scale. GPT and Llama performed similarly, but Mixtral lacks a token for '10'. We therefore selected the 0\,--\,4 scale.} %
The complete score prompt consequently looked as follows:
{
\vspace{-1mm}
\begin{lstlisting}[stringstyle={\textit},escapeinside=!!]
!\textbf{\mbox{User Prompt}}!: "SNIPPET: !\color{gray}\{method\}! Does the code snippet implement the algorithm !\color{gray}\{algorithm\}!?
Score the snippet from '0' to '4', where:
0: The code does not implement the algorithm.
1: !\mbox{\textit{The code shares similarities with the algorithm}}! but most      !\,!likely does not implement the algorithm.
2: !\mbox{\textit{The code appears to implement the algorithm, but may}}! not      !\,!fully match the algorithm's specification.
3: !\mbox{\textit{The code most likely implements the algorithm}}! with minor      !\,!variations.
4: The code implements the algorithm.
Strictly respond with a number from the choices above."
\end{lstlisting}
\vspace{-1mm}
} %
We can then set a \textit{Score Threshold (ST)}, where methods with LLM-assigned scores $\geq$ ST are classified as positive.
By increasing the ST we are able to trade recall for precision.

\subsubsection{Results}\label{sec:rq1:baseline:results}
\begin{figure}[t]
	\centering
	\includegraphics[width=0.45\textwidth, trim=0 0 0 0, clip]{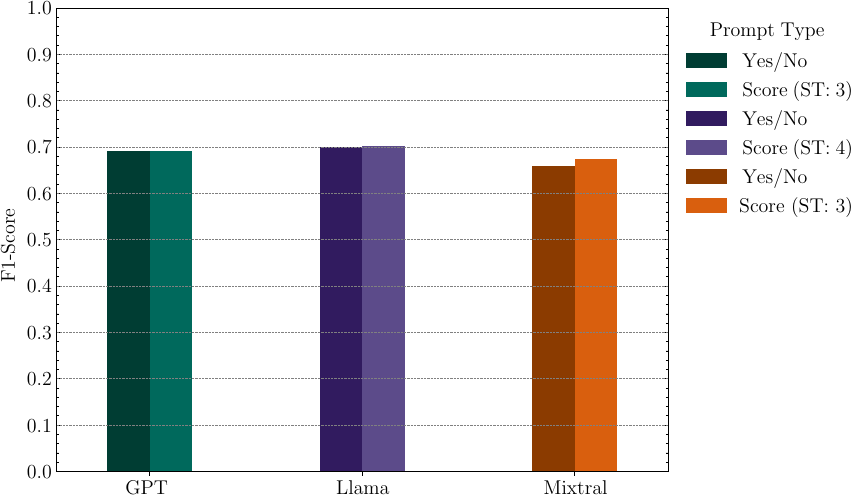}
	\vspace{-1mm}
	\caption{Average F1-score for Yes/No and Score prompting.}
  \vspace{-6mm}
	\label{fig:prestudy_yes_no_vs_score} %
\end{figure}

To choose a sensible baseline for further experiments we compared the Yes/No and the score-prompt.
Since OpenAI's API pricing is based on token count, we used the reduced dataset for GPT.
\Cref{fig:prestudy_yes_no_vs_score} shows the average F1-score over all algorithms in the dataset for each of the evaluated LLMs with the two prompting styles.
When using the score prompt we have to decide which value to choose for the score threshold.
We want to ensure a fair evaluation between the models, however the optimal score threshold may differ depending on the model (and prompting style).
We therefore decided to evaluate all possible score thresholds for each LLM and select the one with the highest F1-score for each configuration individually.
We applied this approach in all following experiments and report the selected score threshold for each configuration.

\begin{figure}[b]
  \vspace{-6mm}
	\centering
	\includegraphics[width=0.35\textwidth, trim=0 0 0 0, clip]{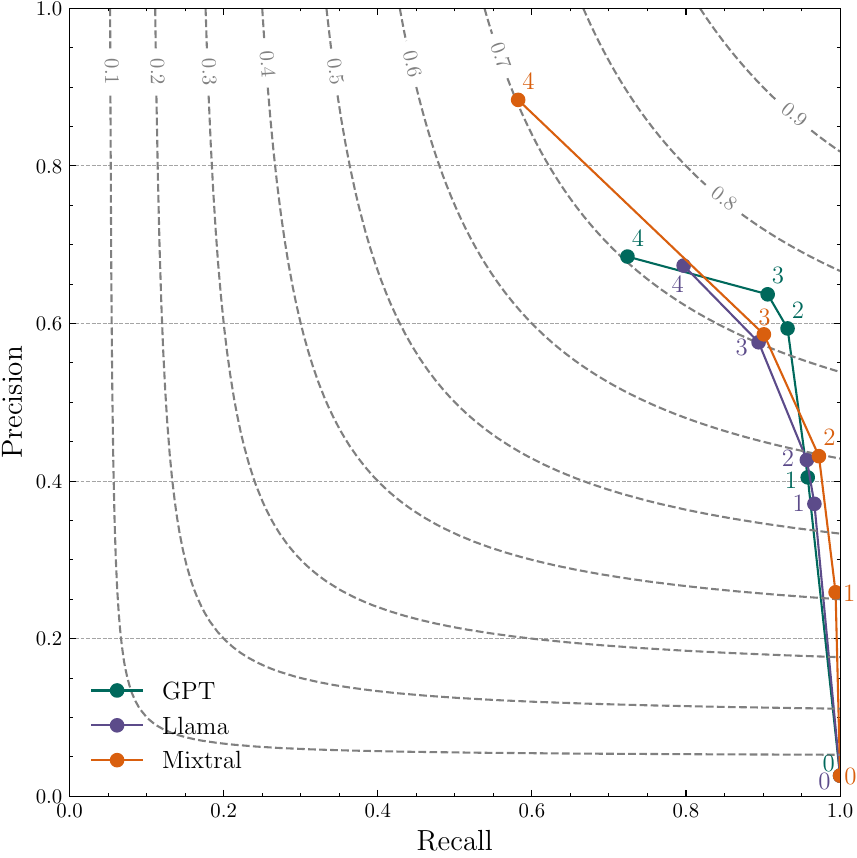}
  \vspace{-1mm}
	\caption{Precision and Recall for different score thresholds.} %
	\label{fig:precision_recall_plot_vs_ST_better}
  \vspace{-1mm}
\end{figure}

As we can see the yes/no as well as the score prompt achieve very similar results.
We can also notice that for all LLMs a relatively high score threshold of 3 or 4 is optimal.
To better illustrate the precision recall tradeoff we also created \cref{fig:precision_recall_plot_vs_ST_better}.
We can see that increasing the score threshold indeed trades off recall vs precision and works especially well in the case of Mixtral.
However, all models have a bias towards recall which is particularly pronounced for Llama and GPT.
Even with a score threshold of 4 their precision is still shy of 0.7.

\subsubsection{Discussion}
In our experiment, the binary (Y/N) and score prompts yield very similar F1-scores.
However, the score prompt offers the added benefit of enabling adjustments to the recall-precision trade-off.
For this reason, we select the score prompt as the baseline for all subsequent experiments.

\subsection{In-Context Learning}\label{sec:In_Context_Learing}
The use of in-context learning (ICL) was popularized by Brown et al.~\cite{Brown2020} and has since become a popular approach for utilizing LLMs~\cite{zhao2023surveyLLMs}.
With ICL (multiple) input-output examples are provided to the LLM in order to guide predictions~\cite{Dong2024}.
One challenge with using ICL is the question of which examples to select.
In our problem setting we can provide demonstrations using the positive or negative class.
I.e. we can use examples of actual algorithm implementations classified as positive or use code examples that do \textit{not} implement the algorithm and are classified as negative. 
Since both types of examples provide information about the expected output it is not clear a priori which would work best for our use case.
We therefore decided to test the following combinations:
\begin{inlistAsWell}
	\item two positive examples (2P+0N)
	\item two negative examples (0P+2N)
	\item two positive examples \textit{and} two negative example (2P+2N)
	\item four positive \textit{and} four negative examples (4P+4N)
\end{inlistAsWell}.
The design of this experiment allows us to evaluate which type of examples are more beneficial, as well as whether providing more demonstrations leads to improved performance. 
Regarding the prompt itself we would use the same prompting template as with the score prompt.
To populate the context we first instantiate the template \( n \in \{2, 4, 8\} \) times to provide the examples with their expected output,
before finally instantiating the template for the method in the dataset we want to classify and invoking the LLM.

\subsubsection{Pre-study Regarding Negative Examples}
We also wondered whether negative examples consisting of random code or code that shares conceptual or structural similarity with the algorithm \dash without actually implementing it \dash would lead to better results. %
We therefore defined the following two type of negative examples:
$(i)$ \textit{random negatives} are methods that share no similarities with the seven algorithm types used in our dataset.
In contrast, $(ii)$ \textit{similar negatives} share conceptual or structural similarity with the algorithm in the dataset without implementing the algorithm itself.
Examples for similar negatives with respect to Bubble Sort are Insertion- or Selection-Sort while similar negatives for Binary Search are Linear- or Ternary-Search.

Both positive and negative examples were retrieved from online coding and Q\&A sites.
Regarding the positive examples we made sure to select reference implementations of the algorithms that cover multiple implementation variants such as iterative and recursive.  
The random negative examples were chosen to strictly not share similarities with the seven algorithm types used in our dataset. %
The similar negative examples were selected to share conceptual or structural similarity for each algorithm and multiple candidates were discussed among three researchers before deciding on the most suited.
To guarantee their functionality and correctness we created unit tests for all selected examples.
The use of the scoring prompt along with in-context learning required us to provide a score for the examples as expected output. %
Therefore, all examples were independently scored by three researchers and disagreements resolved through discussion.

To compare both types of negative examples, we used the (0P+2N) and (2P+2N) configurations to evaluate both the individual influence of negatives and potential interaction effects with the positive examples.
Our pre-study results showed that similar negatives generally yield better performance.
One possible explanation could be that the examples help the model to make a sharper distinction in edge cases where the implementation is from a similar domain but does not fulfill the specification of the algorithm in question. %
This may be because they help the model more sharply differentiate edge cases where implementations are related but do not meet the algorithm's exact specifications.
Accordingly, we chose \textit{similar negative examples} to evaluate them in all combinations and present their results in the next section.

\subsubsection{Results}
\begin{figure}[t]
	\centering
	\includegraphics[width=0.49\textwidth, trim=0 0 0 0, clip]{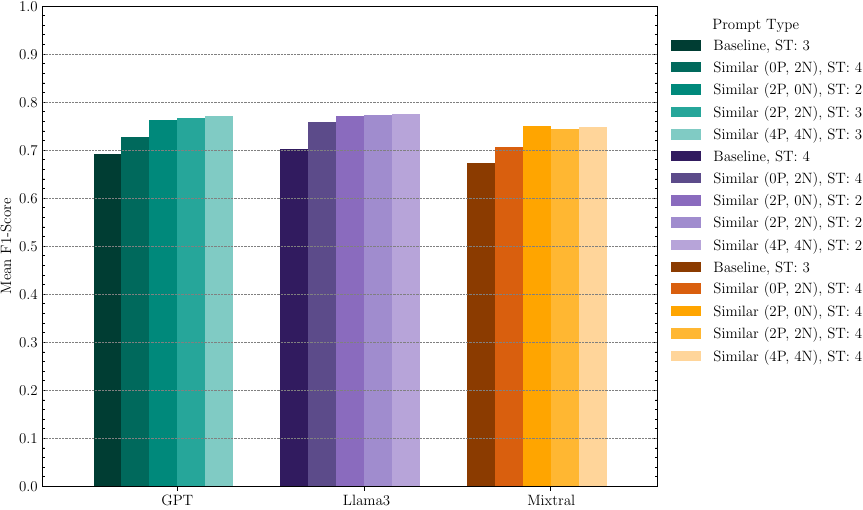}
	\vspace{-5mm}
	\caption{Average F1-score of in-context learning with varying numbers of examples vs the baseline.}
  \vspace{-6mm}
	\label{fig:comparison_score_baseline_and_in_context_learning} %
\end{figure}

\Cref{fig:comparison_score_baseline_and_in_context_learning} displays the F1-score achieved by the LLMs when using the different example combinations compared to the baseline.
For all models providing examples in the context increases performance compared to the baseline.
GPT improves from a baseline F1-score of 69\% to 77\% while Llama improves from 70\% to 78\% with the (4P+4N) combination.
For GPT as well as Llama performance additionally increases consistently when more examples are provided.
While Mixtral has a similar trend the combination with only two positive examples also achieves the highest F1-score with 75\%. 
We can also see that a large share of the improvement for all models can already be achieved by providing only two positive examples.

\Cref{tab:model_comparison_RQ1} shows the F1-score as well as the average inference time per method for the LLMs.
We can see that when more examples are provided in the context the inference time increases linearly.
This is expected since the LLM has to process more tokens when more examples are provided.
With all eight examples, the inference time increases by a factor of $3.\overline{03}$ for Llama and $\sim2.25$ for Mixtral.
While such a difference might not seem like much, the fact that our test-split of BCEval consists of 73,537 methods means that the total inference time increases from 7 and 9 hours to about 20 hours for a single run.

Another interesting insight is that, with the exception of the (4P+4N) combination Mixtral is always slightly slower than Llama, even though it's mixture of experts architecture claims performance improvements as one of its benefits~\cite{Jiang2024MixtralPaperSmallModels}.
One reason for this could be the fact that our scoring prompt requires only the generation a single token which means the LLMs do not have to generate much output-text.
As we will see in \cref{sec:CoT_Discussion} Mixtral inference time is lower compared to Llama when more output text has to be generated.

{
\setlength{\tabcolsep}{4.4pt} %
\begin{table*}[h]
\centering
\small
\begin{adjustbox}{width=0.95\linewidth,center}
\begin{threeparttable}
\begin{tabular}{l|cccc|cccS[table-format=2.2]c|cccS[table-format=2.2]c}
\toprule
\multirow{2}{*}{\textbf{Prompt}}
& \multicolumn{4}{c|}{\textbf{GPT\tnote{1}}} 
& \multicolumn{5}{c|}{\textbf{LLaMa 3}} 
& \multicolumn{5}{c}{\textbf{Mixtral}} \\
& \textbf{ Prec.} & \textbf{Rec.} & \textbf{F1-Score} & \textbf{ST}
& \textbf{ Prec.} & \textbf{Rec.} & \textbf{F1-Score} & \textbf{Inf. Time (s)} & \textbf{ST}
& \textbf{ Prec.} & \textbf{Rec.} & \textbf{F1-Score} & \textbf{Inf. Time (s)} & \textbf{ST} \\
\midrule
Baseline              & 0.64 & 0.91 & 0.69 & 3     & 0.67 & 0.80 & 0.70 & 0.33 & 4     & 0.59 & 0.90 & 0.67 & 0.43 & 3 \\
0P+2N                 & 0.71 & 0.84 & 0.73 & 4     & 0.75 & 0.83 & 0.76 & 0.48 & 4     & 0.66 & 0.86 & 0.71 & 0.55 & 4 \\
2P+0N                 & 0.73 & 0.91 & 0.76 & 2     & 0.72 & 0.92 & 0.77 & 0.48 & 2     & 0.82 & 0.73 & 0.75 & 0.55 & 4 \\
2P+2N                 & 0.77 & 0.86 & 0.77 & 3     & 0.73 & 0.91 & 0.77 & 0.64 & 2     & 0.75 & 0.81 & 0.74 & 0.68 & 4 \\
4P+4N                 & 0.78 & 0.86 & 0.77 & 3     & 0.73 & 0.91 & 0.78 & 1.00 & 2     & 0.81 & 0.77 & 0.75 & 0.96 & 4 \\
CoT\tnote{2}          & 0.59 & 0.93 & 0.68 & 2     & 0.78 & 0.65 & 0.67 & 21.64 & 4    & 0.71 & 0.85 & 0.72 & 6.67 & 3\\
\bottomrule
\end{tabular}
\begin{tablenotes}
  \footnotesize
  \item[] \textsuperscript{1} Due to API costs GPT is evaluated on the reduced dataset. GPT's API does not provide inference times. 
  \item[] \textsuperscript{2} Due to the excessive runtime, CoT is evaluated on the reduced dataset for all models.
\end{tablenotes}
\end{threeparttable}
\end{adjustbox}
\vspace{-1mm}
\caption{Comparison of different metrics across LLMs and prompting techniques. The used score threshold is denoted as ST and the inference time is the average per method.}
\label{tab:model_comparison_RQ1}
\vspace*{\belowTableSaveSpace}
\end{table*}
}

\subsubsection{Discussion}
From the experiment, we conclude that in-context learning improves performance by 4--8 percentage points (pp).
We also find that positive examples have a higher relative improvement in performance compared to negative examples.
Providing more than two positive examples only marginally increases performance while at the same time linearly increasing the inference time.
Therefore, the (2P+0N) combination seems like a \enquote{sweet-spot} for our use case, offering the highest relative improvement with minimal overhead.

\subsection{Chain of Thought Prompting}
Chain of thought (CoT) is a prompting strategy that can significantly improve performance in a variety of complex reasoning tasks~\cite{Wei2022, Kojima2022, zhao2023surveyLLMs}.
The main idea behind CoT consists of instructing the LLM to produce a step by step explanation or reasoning before the final answer.
While CoT can be combined with ICL~\cite{Zhang2022}, we purposefully decided to evaluate both separately in order to be able to identify their individual influence on the performance.
Moreover, Kojima et al.~\cite{Kojima2022} show that CoT without examples performs quite well, since LLMs are already able to generate reasoning chains due to their pretraining.

To use CoT we replace the last sentence of the score baseline prompt explained in \cref{sec:baseline} with the following instruction and allow the LLM to create up to 300 tokens for the explanation:
{
\vspace{-1mm}
\begin{lstlisting}[stringstyle={\textit},escapeinside=!!]
!\mbox{Lets think step-by-step, then answer with a number} from the choices above.
\end{lstlisting}
\vspace{-1mm}
}

In the previous experiments we only expected the LLM to generate a single token as output.
Therefore, we could use the logarithmic probabilities (logprobs) for each token directly to select the most probable token.
This has the benefit of avoiding the influence of parameters like temperature~\cite{Ackley1985, Ficler2017}, top-k~\cite{Fan2018} and top-p (i.e., Nucleus Sampling~\cite{Holtzman2020}) and is a common technique for evaluating LLMs also used with the well-known MMLU dataset~\cite{Fourrier2023}.
With CoT however we want the LLM to generate multiple tokens, which means that the choice of parameter becomes relevant.
For the CoT experiment we want to allow the model some slight exploration when generating the reasoning while still ensuring accuracy rather than diversity of model responses.
We therefore selected a moderate temperature of 0.5 to balance both aspects.
Additionally, we used the standard values of 50 and 0.9 for top-k and top-p, which moderately constrain the output space while still allowing for a degree of variability.
Generating the explanations incurs a heavy increase in inference time, which makes running the CoT prompt infeasible on the whole dataset.
We therefore decided to only evaluate the CoT prompting on the test split of the reduced dataset.

\subsubsection{Results}
\begin{figure}[t]
	\centering
	\includegraphics[width=0.49\textwidth, trim=0 0 0 0, clip]{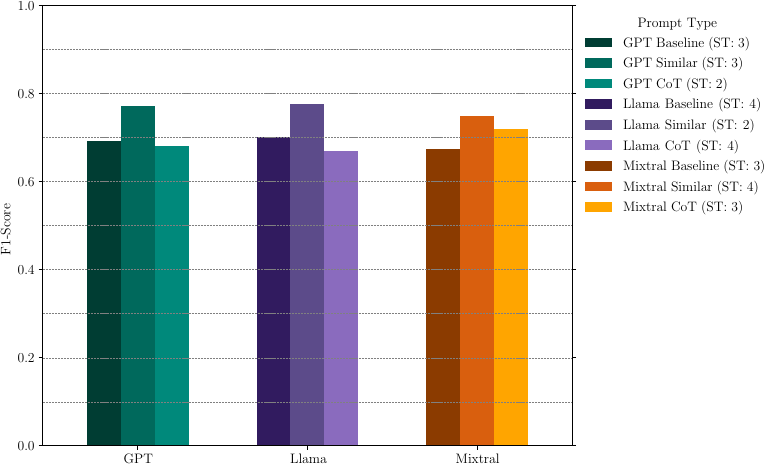}
	\vspace{-6mm}
	\caption{Average F1-scores for the Baseline, (4P+4N) as the best performing ICL combination and CoT.}
  \vspace{-6mm}
	\label{fig:comparison_CoT_baseline_ICL} %
\end{figure}

\Cref{fig:comparison_CoT_baseline_ICL} displays the results for CoT compared to the baseline and the best performing in-contest learning combination (4P+4N) from our previous experiments.
With CoT prompting GPT, Llama and Mixtral achieved F1-scores of 68\%, 67\% and 72\% respectively.
Surprisingly the use of CoT prompting leads to a decrease in performance for the GPT and Llama models when compared to their baselines.
Wile Mixtral does benefit from CoT prompting the improvement is not as pronounced as with in-context learning.

Regarding the runtime, the average inference time per method are 21.64s for Llama and 6.67s for Mixtral.
This is an increase by a factor of $65.\overline{57}$ and 15.51 compared to the baseline.
Running CoT prompting on the full test split would thus require approximately 19 and 6 days.

\subsubsection{Discussion}\label{sec:CoT_Discussion}
We find that CoT prompting is inferior when compared to ICL, both in terms of achieved F1-score as well as runtime.
One possible explanation for this could be that our models are too small to take full advantage of CoT.
This explanation is supported by the experiments of Wei et al.~\cite{Wei2022} who find that CoT prompting only shows its positive effects on larger size models.
However, Llama and Mixtral are quite large and require two H100 GPUs with a combined 188GB of VRAM to execute a single model.
Another interesting observation is the fact that with the increased text generation Mixtral becomes faster than Llama.
This indicates that the runtime benefits of the sparse mixture-of-experts architecture only show with tasks that require larger amounts of text generation. %

\begin{AnswerBox2}
  \textbf{Answer to \ref{rq:1}:}~\textit{
  Our first experiment shows that score prompting performs comparably to Yes/N prompting, with the added benefit of being able to trade recall for precision.
  In-context learning (ICL) improves performance by 4--8~pp, with positive examples leading to a larger improvement than negative ones. 
  While providing more than two positive examples leads to a slight improvement they also increase the inference time proportionally.
  Chain-of-thought prompting did not surpass ICL for any LLM.  %
  Therefore, using ICL with two positive examples seems to be a \enquote{sweet-spot}, reaping most of the improvements with only a slight increase of runtime. 
  }
\end{AnswerBox2}

\section{\ref{rq:2}: Combination with Static Analysis}\label{sec:RQ2} %
In \cref{sec:RQ1}, we already noticed that more complex prompts lead to better results but also increase the runtime of an LLM-based algorithm detection approach.
Additionally, both the size of LLMs as well as the size of source code bases continue to increase, further negatively influencing the runtime.
In \ref{rq:2}, we therefore focus on improving the runtime of our detection pipeline by using filter-patterns designed to exclude methods that are unlikely to implement the algorithm in question.
The excluded methods are directly assigned a to the negative class and not shown to the LLM.
Therefore, recall is comparatively more important for our filter-patterns while the precision can be increased in the subsequent classification by the LLMs.

We evaluate two pattern types: \textit{Keyword-Based} and \textit{Structural} filter patterns. %
As the name suggest the former is designed to match typical keywords used in implementations such as identifier or method names.
These patterns are based on regular expressions (regex) that are matched on the source code text.

The structural filter patterns are based on the program AST.
Using the DSL developed by Neumüller et al.~\cite{Neumueller2024} we describe a minimal set of features a method must possess to be considered a match and passed to the LLM.
E.g. in the case of Bubble Sort we require two nested loops containing an if-statement with the assignment of an array read.
These patterns are matched by statically analyzing the source code and performing graph matching between the filter patterns and the program AST.

Besides precision, recall, and F1-score, we compute the average reduction in methods passed to the LLM for evaluation.
This enables pattern comparison independent of runtime.

To avoid bias and ensure robustness, we created our patterns using eight example implementations per algorithm, which we retrieved from online coding and Q\&A sites.
We finetune and compare the patterns on the test-split before reporting the best patterns performance for each category (\textit{Keyword-Based} and \textit{Structural}) on the validation split to demonstrate generalizability and avoid overfitting
\Cref{tab:filter_pattern_comparison_RQ2} shows the results of all filter patterns in combination with the LLMs.
We evaluate the filter patterns using the baseline prompting style and configuration from~\cref{sec:baseline}.

\subsection{Keyword-Based Patterns}
Developers often use self-descriptive identifiers and follow style guides to improve readability.
A large-scale study of Java projects by Zhang et al.~\cite{Zhang2020a} found that identifiers commonly include nouns, verbs, and adjectives.
Building on these insights we create two types of keyword-based patterns, which we call \textit{\enquote{Recall Focused}} and \textit{\enquote{Recall Focused Enhanced Precision}}.

\subsubsection{Recall Focused}
These patterns aim to maximize recall to avoid excluding any true positives.

\underline{Creation recipe:}
First, we extracted all identifiers from the example implementations. 
Next we divided identifiers into explicit (e.g., \textit{transposeMatrix, transpose}) and generic (e.g., \textit{rows, cols, temp, \ldots}). 
We then defined regular expressions for each group and introduced a hit threshold that specifies how many keywords must be found at a minimum.
Explicit keywords use a hit threshold of 1, meaning that only one of these keywords needs to be matched.
Regarding the more generic identifiers, multiple (2--3) keywords must co-occur for the method not to be excluded.
The hit-thresholds were determined empirically on the test set.
For Bubble Sort our explicit keywords are \textit{bubble} and \textit{sort}, while the second group consists of the following keywords with a threshold of 3:
\textit{arr, temp, i, j, size, input, list, length}.

\underline{Results:}
\Cref{tab:filter_pattern_comparison_RQ2} shows the standalone performance of the filter patterns and LLMs as well as their combination.
Looking at the standalone performance of the \textit{{Recall Focused}} patterns we can see that we achieved our goal of high recall retraining 96\% of all implementations, 
while at the same time achieving a good reduction of 72.39\%.
As these patterns are designed for the combination with LLMs their precision is quite low, which of course also leads to a low F1-Score.

Combining them with the LLMs has minimal impact on classification performance, as only 18 of 1,041 true positives are missed across all algorithms, slightly reducing recall.
The impact of the missed implementations varies slightly across the different LLMs, as it also depends on how each LLM classifies the methods. 
Excluding an algorithm implementation the LLM would have classified as a false negative has no effect, while excluding one the LLM would have classified as a true positive reduces recall.
Overall, the pattern works well, especially considering that it is relatively simple to construct. 
However, the use of many generic keywords in the second category means that the LLMs are still prompted quite frequently.
In the next section we therefore create a second set of keyword-based patterns with a stronger focus on precision.

{
\setlength{\tabcolsep}{3.5pt} %
\begin{table*}[h]
\centering
\small
\begin{adjustbox}{width=0.95\linewidth,center}
\begin{threeparttable}
\begin{tabular}{l|ccc|cccc|cccc|cccc|S[table-format=1.2]}
\toprule
\multirow{2}{*}{\diagbox[width=3.5cm]{\textbf{Filter Pattern}}{\textbf{LLM}}}
& \multicolumn{3}{c|}{\textbf{None}}  
& \multicolumn{4}{c|}{\textbf{GPT\tnote{1}}} 
& \multicolumn{4}{c|}{\textbf{LLaMa 3}} 
& \multicolumn{4}{c|}{\textbf{Mixtral}} 
& \multicolumn{1}{c}{\multirow{2}{*}{\textbf{Reduction (\%)}}} \\
& \textbf{Prec.} & \textbf{Rec.} & \textbf{F1}
& \textbf{Prec.} & \textbf{Rec.} & \textbf{F1} & \textbf{ST}
& \textbf{Prec.} & \textbf{Rec.} & \textbf{F1} & \textbf{ST}
& \textbf{Prec.} & \textbf{Rec.} & \textbf{F1} & \textbf{ST} \\

\midrule
None (Baseline)                                                   & n/a & n/a & n/a        & 0.60 & 0.92 & 0.68 & 3     & 0.67 & 0.80 & 0.70 & 4     & 0.59 & 0.90 & 0.67 & 3     & n/a \\
Recall Focused                                                    & 0.13 & 0.96 & 0.20     & 0.60 & 0.89 & 0.68 & 3     & 0.68 & 0.76 & 0.70 & 4     & 0.61 & 0.88 & 0.69 & 3     & 72.39\% \\
Recall Focused Enh. Prec.                                         & 0.28 & 0.88 & 0.40     & 0.63 & 0.82 & 0.68 & 3     & 0.71 & 0.71 & 0.69 & 4     & 0.65 & 0.82 & 0.70 & 3     & 89.80\% \\
Prominent Feature                                                 & 0.71 & 0.79 & 0.71     & 0.81 & 0.73 & 0.73 & 2     & 0.76 & 0.76 & 0.73 & 1     & 0.72 & 0.78 & 0.72 & 1     & 97.12\% \\
Neumüller et al.~\cite{Neumueller2024}\tnote{2}                   & 0.77 & 0.76 & 0.74     & 0.87 & 0.75 & 0.79 & 3     & 0.85 & 0.76 & 0.78 & 2     & 0.87 & 0.75 & 0.79 & 3     & 97.50\% \\
\midrule
\multicolumn{17}{c}{\textbf{Validation Set Results}} \\
\midrule
None (Baseline)                                                   & n/a & n/a & n/a        & 0.60 & 0.91 & 0.69 & 3     & 0.72 & 0.85 & 0.75 & 4     & 0.87 & 0.69 & 0.72 & 4     & n/a \\
Recall Focused Enh. Prec.                                         & 0.31 & 0.88 & 0.43     & 0.73 & 0.82 & 0.74 & 3     & 0.78 & 0.76 & 0.74 & 4     & 0.72 & 0.83 & 0.75 & 3     & 90.22\% \\
Neumüller et al.~\cite{Neumueller2024}\tnote{2}                   & 0.78 & 0.81 & 0.76     & 0.85 & 0.81 & 0.81 & 3     & 0.85 & 0.81 & 0.81 & 1     & 0.86 & 0.81 & 0.81 & 3     & 97.46\% \\
\bottomrule
\end{tabular}
\begin{tablenotes}
  \footnotesize
  \item[] \textsuperscript{1} GPT evaluated on the full dataset. \textsuperscript{2} Results for \Standalone exclude Transpose Matrix, as they did not publish a pattern this algorithm.
\end{tablenotes}
\end{threeparttable}
\end{adjustbox}
\vspace{-1mm}
\caption{Performance comparison of different filter patterns across LLMs on training and validation sets. ST indicates the score threshold used for classification, while Reduction(\%) denotes the proportion of excluded methods relative to the baseline.}
\label{tab:filter_pattern_comparison_RQ2}
\vspace*{\belowTableSaveSpace}
\end{table*}
}

\subsubsection{Recall Focused Enhanced Precision}
With these patterns, our goal was to increase precision considerably while maintaining high recall.
Although there is no silver bullet that works for each algorithm, the following modifications proved effective in achieving this.

\underline{Creation recipe:}
We removed overly generic keywords shared across multiple patterns and lowered the threshold, to account for the smaller keyword set.
We also merged explicit and generic keywords into a single category with higher hit threshold, which was effective for Bubble Sort and Transpose Matrix.
Furthermore, we combined frequently co-occurring identifiers into a single regular expression to ensure they appear together for a match.
As an example the pattern for Binary Search consisted of two groups: the first required a single match from the explicit keywords \textit{binary} or \textit{search}, while the second required at least two matches from the following regular expressions: \textit{mid, sorted, low.*high, min.*max, first.*last, start.*end}.

\underline{Results:}
Evaluating the pattern in isolation we can see that our changes moderately improve their precision to 28\%,
while at the same time achieving a substantial reduction of 89.80\%, as shown in \cref{tab:filter_pattern_comparison_RQ2}.
However, these changes also lead to the exclusion of 131 out of 1,041 true positive implementations.

Combining the patterns with the LLMs therefore also noticeably decreases recall for all three models.
Interestingly, starting with the \textit{Recall Focused Enhanced Prec.} patterns, combining LLMs with filters always yields higher precision than either approach in isolation.
We attribute this to the filter patterns removing many methods the LLMs might otherwise misclassify as false positives.
This aligns with our findings from \cref{sec:rq1:baseline:results}, where \cref{fig:precision_recall_plot_vs_ST_better} illustrates that the LLMs tend to favor recall over precision.
Our patterns eliminate many true negatives, leaving the LLMs fewer opportunities to incorrectly classify these as false positives.

\subsection{Structural Patterns}
Our structural patterns are based on describing key structural features typical algorithm implementations possess.
To achieve this we use the domain specific language (DSL) developed by Neumüller et al.~\cite{Neumueller2024} which is designed to capture the key features of an algorithm and can be used to express a search-pattern on the abstract syntax tree.
Using static code analysis and graph matching, patterns are compared to the implementation, and only matching implementations are passed to the LLMs.
In the following we compare two kinds of pattern: \textit{Prominent Feature} and \Standalone pattern.

\subsubsection{Prominent Feature}
The objective of the \textit{Prominent Feature} patterns is to further enhance precision compared to the \textit{keyword-based} patterns, while preserving high recall.
These patterns capture only the most important features that are characteristic for the implementations of a specific algorithm.

\underline{Creation recipe:}
We examined our set of example implementations and identified recurring operations and structures across multiple implementation variants of the same algorithm.
Examples of key features regarding Bubble Sort include a nested loop with array access, whereas Binary Search is characterized by a mid-point calculation in combination with iteration or recursion.
A key feature of Transpose Matrix implementations is an assignment consisting of an array-write and an array-read using the same indices, but in reversed order.
We then expressed only these prominent key-features using the DSL to create the filter patterns.
Using such light-weight features also makes it possible to find implementations that are embedded as part of larger methods, where more complex and fully fledged search-patterns might fail due to the surrounding code and possible implementation details.
This is especially important for algorithms like Transpose Matrix or Binary Search that are often implemented as part of larger more complex methods.

\underline{Results:} %
Compared to the \textit{keyword-based} patterns the \textit{Prominent Feature} patterns achieve a significantly higher precision (71\%) and reduction rate (97.12\%).
Only 99 true positive implementations are excluded, which is fewer than the 131 missed by the \textit{Recall Focused Enhanced Prec.} patterns.
Interestingly, both \textit{structural} patterns are able to outperform the LLM-only baseline in terms of precision and F1-Score.
However, they also require more manual effort in their creation compared to using the LLMs.

Combining the \textit{Prominent Feature} pattern with the LLMs further increases precision without harming recall.
Importantly this results in better precision and F1-Scores than either method achieves in isolation.
Due to their strong reduction rate combining these patterns with the LLMs also leads to further decreases in runtime. 
Given an average inference time of 0.96 seconds per method with Mixtrals best performing prompt from \cref{sec:In_Context_Learing}, BCEval's total processing time decreases from 19,3 hours to just 33 minutes.

We also observe a change in the score thresholds for all models.
As more true negatives are removed by pre-filtering — reducing the risk of misclassification — lower thresholds become viable, despite the LLMs recall bias.
This suggests that the patterns effectively remove many methods that LLMs might otherwise misclassify, improving the precision-recall trade-off.
Additionally, the distinction between the best and second-best similarity thresholds typically narrows to less than 1 percentage point.
It is therefore important to not over-interpret slight changes in score thresholds if pre-filtering is used.

\subsubsection{\Standalone}
To evaluate their DSL, Neumüller et al.~\cite{Neumueller2024} also published a set of algorithm search-patterns for BCEval.
Unlike our \textit{Prominent Feature} patterns — which are lightweight filter heuristics to use with LLMs — their patterns are standalone solutions targeting both high precision and high recall by themselves.
As a result, these patterns are more extensive and describe a wider range of algorithmic features and implementation variants. 
Consequently, they are also significantly more time-consuming to construct.

\underline{Creation recipe:}
Compared to our \textit{Prominent Feature} patterns, Neumüller et al. describe further details regarding the structure and operations of the algorithms.
For example, their pattern for Binary Search additionally describes a fully specified conditional branch that compares the search-key with the current array element. 
The pattern also covers implementations that use the right-shift operation to calculate the midpoint.
Similarly, the Bubble Sort pattern additionally describes all the assignments that are used to perform the swapping of unsorted elements.

\underline{Results:}
As the search-patterns published by Neumüller et al.~\cite{Neumueller2024} do not include a specification for Transpose Matrix, we report results for their patterns excluding this algorithm.
The \Standalone patterns achieve the best precision and F1-Score among all patterns.
However, compared to the \textit{Prominent Feature} patterns, they only yield a marginal improvement in method reduction (97.50\% vs. 97.12\%).
It is worth noting that 239 true positive implementations are excluded across all algorithms, which is the highest among all evaluated filter patterns and a result of their focus on precision and recall as standalone patterns.
Interestingly, most (218) of the missed true positives originate from a single algorithm which is Binary Search.
This indicates that the Binary Search pattern is too restrictive and likely fails to account for a specific implementation variant.

When combined with the LLMs their strong performance results in the highest precision and F1-scores for each model.
As with the \textit{Prominent Feature} patterns, combining the \Standalone patterns with LLM classification outperforms either technique in isolation, with respect to precision and F1-score.

\subsection{Validation} %
To assess robustness and generalizability, we evaluate the best-performing keyword and structural patterns, LLMs, and their combination on the validation split.
Evaluating the patterns and LLMs separately shows that both perform slightly better on the validation set.
F1-Score differences range from 2--3 percentage points (pp) for the patterns to 1--5~pp for the LLMs~(cf. Table~\ref{tab:filter_pattern_comparison_RQ2}).
This is most likely due to the smaller sample size and potential sampling variance within the validation set.

Comparing the F1-scores of the \textit{Recall Focused Enhanced Prec.} pattern and LLM combinations between splits, only GPT exhibits a difference exceeding 5~pp. 
For the \Standalone pattern, all differences in F1-scores are consistently under 5~pp. 
Importantly, the relative ranking of the patterns also remains stable across the splits. 
In summary, the filtering patterns exhibit comparable performance on the validation set, substantially reducing the number of methods while simultaneously improving classification performance.
Our overall conclusions for RQ2 are as follows:

\begin{AnswerBox2}
  \textbf{Answer to \ref{rq:2}:}~\textit{The combination of static analysis in the form of filter patterns with LLMs is surprisingly effective for decreasing runtime while also improving classification performance.
  If runtime reduction is the main goal, the \emph{Recall Focused} patterns offer a strong trade-off: they are simple to construct and achieve substantial method reduction (72.39\%) with minimal impact on LLM classification (F1-scores remain stable in the 68--70\% range).
  The \emph{Recall Focused Enhanced Prec.} and \emph{Prominent Feature} patterns require some manual tuning for each algorithm but yield significantly better reduction rates (89.80--97.12\%) and improved F1-scores (up to 73\%). 
  Furthermore, starting with the \emph{Recall Focused Enhanced Prec.} pattern the combination of filter patterns and LLMs outperforms each individual approach in terms of precision and F1-score.
  Lastly, the \Standalone patterns offer the best classification results overall but demand substantial manual effort. 
  Their marginal gain in method reduction over the \emph{Prominent Feature} patterns (97.50\% vs. 97.12\%) is offset by a notable increase in precision and F1-scores as high as 81\%.
  Therefore, they are most suitable when maximizing classification performance is paramount.}
\end{AnswerBox2}

\section{\ref{rq:3} Name Information} \label{sec:RQ3}
In this section, we quantitatively assess the extent to which LLMs rely on name-information contained in the code for their classification.
This is important since an over-reliance on name-information, would limit their ability to correctly classify implementations that are not using indicative identifier names.
We are further motivated by our results in \cref{sec:RQ2}, where we find that simple keyword-based search-patterns can effectively filter many false positives while maintaining high recall.
This indicates that the naming-information in the code can be used for classification, at least to some degree.

it is also well established that machine learning models can place too much attention on superficial features like identifier names instead of the semantics of the code.
For instance, Chourasia et al.~\cite{Chourasia2022} report that their CodeBERT-based classifier for programming assignments did exactly this before they adapted the training process using obfuscation to avoid such behavior.
Given that the training details of most LLMs are not publicly disclosed, investigating this potential over-reliance on name information becomes especially important.

To this end, we conduct an ablation study in which we systematically remove name information from the source code.
We repeat the baseline experiment using the same settings (including the score threshold), but obfuscate identifiers in the source code by assigning them random names.

\begin{table*}[h]
\centering
\small
\begin{adjustbox}{width=0.95\linewidth,center}
\begin{threeparttable}
\begin{tabular}{l|cccc|cccc|cccc}
\toprule
\multirow{2}{*}{\textbf{Prompt}} 
& \multicolumn{4}{c|}{\textbf{GPT\tnote{1}}}
& \multicolumn{4}{c|}{\textbf{Llama 3}}
& \multicolumn{4}{c}{\textbf{Mixtral}} \\
& \textbf{Prec.} & \textbf{Rec.} & \textbf{F1-Score} & \textbf{ST}
& \textbf{Prec.} & \textbf{Rec.} & \textbf{F1-Score} & \textbf{ST}
& \textbf{Prec.} & \textbf{Rec.} & \textbf{F1-Score} & \textbf{ST} \\

\midrule
Baseline                                          & 0.60 & 0.92 & 0.68 & 3     & 0.67 & 0.80 & 0.70 & 4     & 0.59 & 0.90 & 0.67 & 3 \\
Baseline with Obfuscation                         & 0.65 & 0.88 & 0.71 & 3     & 0.79 & 0.84 & 0.79 & 4     & 0.76 & 0.85 & 0.77 & 3 \\
\bottomrule
\end{tabular}
\begin{tablenotes}
  \footnotesize
  \item[] \textsuperscript{1} GPT is evaluated on the full dataset.
\end{tablenotes}
\end{threeparttable}
\end{adjustbox}
\vspace{-1mm}
\caption{Comparison of LLM performance with and without name obfuscation. ST denotes the score threshold.}
\label{tab:obfuscation_comparison_RQ3}
\vspace*{\belowTableSaveSpace}
\end{table*}

\textbf{Results:} \Cref{tab:obfuscation_comparison_RQ3} shows that all models experience a decrease in recall ranging from 4 to 5 percentage points (pp).
This is to be expected since the missing name-information makes the classification more challenging.
Conversely, model precision increases by 5~pp for GPT, 12~pp for Llama, and 17~pp for Mixtral.
Surprisingly the overall F1-Score actually increases by \mbox{3--10~pp} across the models.
One possible explanation is that overly generic or even misleading identifier names, such as a method named \enquote{sort} that does not implement BubbleSort, can no longer lead to misclassifications thereby improving precision. %
At the same time, the LLMs are still able to recognize a large portion of the implementations despite the obfuscation.
Together, these effects could explain the observed improvement in F1-score.

Among the models, Mixtral is most affected, followed by Llama, indicating a comparatively higher dependence on naming cues.
GPT is comparatively less influenced with none of its metrics changing by more than 5~pp. 

\begin{AnswerBox2}
  \textbf{Answer to \ref{rq:3}:}~\textit{
    Obfuscating name information in source code reduces recall by 4--5~percentage points (pp) while simultaneously improving precision.
    Interestingly the overall F1-Score actually increases by 3--10~pp.
    The results indicate that while LLMs do leverage name information, they are still able to recognize most algorithm implementations in the absence of naming cues.
    Since the LLMs appear to also utilize the underlying semantics of the code, they offer a promising avenue for future research on robust algorithm detection.}
\end{AnswerBox2}

\section{Threats to Validity}
  Regarding \textbf{construct validity}, we evaluate our experiments using the macro-averaged F1-score across all algorithms.
  We selected this metric because the number of implementations varies significantly between different algorithms in the dataset (cf.~\cref{tab:dataset_overview}).
  A \textit{micro}-averaged score would therefore place unwarranted importance on algorithms with many instances.
  The macro-averaged F1-score avoids this bias and aligns better with our goal of assessing the general algorithm recognition capability of LLMs across different algorithms.
  
  Concerning \textbf{internal validity}, we base our evaluation on a subset of BCEval~\cite{svajlenko2016bigcloneeval}, which was manually labeled by three judges in over 216 hours of validation efforts and is widely used with over 110 citations according to IEEE Xplore.
  While Krinke et al.~\cite{Krinke2022} consider BCEval valuable for evaluating code clone detection tools they also criticize its use for learning code similarity and the flawed assumptions some approaches make about its clone-\textit{pair} information.
  We however, do not train any machine-learning models and do not make use of, or any assumptions about, the clone-\textit{pair} information.

  To extract and process methods from the dataset, we employ Spoon's~\cite{Spoon} DefaultJavaPrettyPrinter, which may reorder Java modifiers and remove/insert empty lines in certain edge cases.
  Where feasible and conflict-free, fully qualified names are also shortened through imports. 
  As these modifications are purely syntactic, we do not believe that they pose a significant threat.

  Pertaining to \textbf{external validity}, BCEval was created based on real-world source code taken from open-source projects.
  The main threat is the fact that BCEval only consists of algorithms implemented within a single method.
  The algorithms themselves however, are quite diverse including searching, sorting, factorization, and recursion.
  Lastly, BCEval contains only Java code, which is another limitation. 
  However, Java is widely used and a good representative for object-oriented languages.  

  One threat to \textbf{conclusion validity} is the non-deterministic nature of LLMs.
  To mitigate this, we follow best practices by using fixed model snapshots and random seeds, setting a moderate temperature, and evaluating the LLMs via their log probabilities rather than raw text outputs. 
  Furthermore, we observe consistent results across the different LLMs used in our evaluation.

\section{Related Work} %
We establish context by reviewing work on program concept recognition and LLMs, especially their use in code-related tasks.

\textbf{Program Concept Recognition:}
Several authors have worked on identifying concepts in source code~\cite{Kozaczynski1992, Quilici1994, Wills1994, Nunez2017, ARCC2}, including coding techniques, data structures, and architectural patterns,
with some approaches specifically designed for algorithm recognition~\cite{Metzger2000, Alias2003, Zhu2011, Taherkhani2013, Mesnard2016, Nunez2017, ARCC2, Long2022, Neumueller2024}.
Approaches in this domain are often based on graph-matching between search-patterns and code representations (e.g. abstract syntax trees, data- or control-flow graphs) or employ machine learning classifiers (e.g. SVM, deep neural networks) trained on these representations.
Of the few approaches that perform an evaluation, most use benchmarks based on programming contests involving few algorithms and a small negative class with limited representativeness of real-world source code.
Furthermore, ML-based approaches also lack flexibility, as supporting additional algorithms demands significant retraining effort.
Our work improves on both aspects and is the first to comprehensively evaluate the combination of LLMs with static analysis for the use case of algorithm recognition.

\textbf{Large Language Models:}
Many efforts aim to enhance LLMs via improved architectures (attention~\cite{vaswani2023attentionneed}, mixture of experts \cite{mixtral2024BlogMixtral8x22b}), refined prompting (ICL~\cite{Brown2020}, CoT~\cite{Wei2022}), and optimization methods (quantization~\cite{Dettmers2022}, distillation~\cite{sanh2020distilbertdistilledversionbert}) to improve both output quality and runtime efficiency.
Additionally, numerous studies investigate LLMs in code related tasks such as code translation~\cite{Yang2024, Zhang2025}, bug fixing~\cite{2023APR_LLMs, 2023RepairIsNearlyGeneration} or code clone detection~\cite{Dou2023, Khajezade2024}.
For example Khajezade et al.~\cite{Khajezade2024} explore using GPT for code clone detection in both mono-lingual and cross-lingual settings finding that it achieves F1-scores of 87.8\% and 87.7\% respectively, outperforming baselines in the cross-language scenario.
Regarding the area of bug fixing Xia et al.~\cite{Xia2023} find that directly applying state-of-the-art LLMs can already substantially outperform state-of-the-art tools, with Codex fixing 99 out of 255 bugs.
While our setting and focus on algorithm recognition is unique, we share the context of code-related tasks and therefore rely on these earlier works to select the most promising models, prompts and inference parameters.

Lastly, some approaches also investigate the combination of LLMs with static code analysis.
Liu et al.~\cite{Liu2024} demonstrate that augmenting LLM prompts with dependency information obtained via static analysis leads to a performance gain of up to 21.30 percentage points on code generation tasks.
Similarly, Lomshakov et al.~\cite{Lomshakov2024} find that complementing LLM context with call graph information improves code summaries and reduces the number of hallucinations.
Our approach focuses on the distinct task of algorithm recognition and integrates traditional static analysis techniques \textit{before} applying LLMs instead of augmenting the prompt,
while combining the strengths of classical algorithm recognition techniques with LLMs.

\section{Conclusions and Future Work}
This study presents the first comprehensive evaluation of algorithm recognition using LLMs in isolation, as well as in combination with static code analysis, to improve runtime efficiency and classification accuracy.
Regarding prompting strategies, we find that in-context learning works best for our use case.
Compared to the baseline F1-scores of 67--70\%, providing two positive examples improves the scores by about 7--8 percentage points (pp), with only a small increase in inference times. 
Adding more examples (e.g., four positive and four negative) yields only marginal additional gains (1--2~pp), but more than doubles the inference time.
We therefore conclude that two examples provide a \enquote{sweet-spot} balancing classification performance and efficiency.

We then combine LLMs with static analysis techniques to pre-filter unlikely algorithm implementations.
Our findings show that filter patterns significantly reduce LLM requests (by 72.39\% to 97.50\%), while also improving classification performance.
E.g. combining LLMs with the \textit{Prominent Feature} patterns — which are moderately easy to create — simultaneously increases precision (by 9--21~pp) and F1-scores (by 3--5~pp) across models.
Crucially, the combined approach outperforms either technique in isolation, with respect to precision and F1-score.
Lastly, we investigate the reliance of LLMs on name-information in the source code, finding that while LLMs do use this information to some extent, they can still accurately identify the majority of algorithm implementations even when names are obfuscated.

Our overall conclusion is that LLM-based approaches to algorithm recognition are promising and their combination with static analysis is surprisingly effective for decreasing runtime requirements while also improving classification performance.

Further research is needed to fully assess LLM-based algorithm recognition, particularly for complex algorithms.
One interesting direction is the development of an automated pipeline that collects algorithm implementations from public repositories and derives both filter patterns and examples for in-context learning.
Additionally, we envision an IDE plugin preloaded with patterns from a library, enabling developers to perform algorithm recognition with little manual setup.

\section{Data Availability}
Our replication package is available on Zenodo~\cite{neumueller2025CombiningZenodo}.

\ifanonymous %
\else
  \section{Acknowledgments}
   The authors acknowledge support by the state of Baden-Württemberg through bwHPC. 
\fi

\bibliographystyle{IEEEtran}
\bibliography{references}
\end{document}